# Multi-Transmission Node DER Aggregation: Chance-Constrained Unit Commitment with Bounded Hetero-Dimensional Mixture Model for Uncertain Distribution Factors

Weilun Wang, *Graduate Student Member, IEEE*, Zhentong Shao, *Member, IEEE*, Yikui, Liu, *Member, IEEE*, Brent Eldridge, *Member, IEEE*, and Abhishek Somani, *Member, IEEE*, Jesse T. Holzer, Lei Wu, *Fellow, IEEE*

*Abstract* — To facilitate the integration of distributed energy resources (DERs) into the wholesale market while maintaining the tractability of associated market operation tools such as unit commitment (UC), existing DER aggregation (DERA) studies usually consider that each DERA is presented on a single node of the transmission network. Nevertheless, the increasing scale and geographical distribution of DERAs covering multiple transmission nodes, posing new challenges in modeling such multi-transmission-node DERAs (M-DERAs). Indeed, assessing the aggregated impact of an M-DERA on power flows is a non-trivial task, because the sensitivities of each transmission line to DERs at different transmission nodes are not identical. Inspired by the distribution factor (DF) based shift factor (SF) aggregation strategy in industry practice, this paper proposes a novel DF-based chance-constrained UC (CCUC) model to determine system optimal operation plans with M-DERAs. DFs, treated as uncertain parameters to describe possible responses of DERs against aggregated dispatch instructions from regional transmission organizations, are modeled via a bounded hetero-dimensional mixture model (BHMM) by leveraging historical DF records distributed on multiple hyperplanes in a bounded space. With this, power flow limits are modeled as chance constraints in CCUC, which is reformulated into a scenarios-based stochastic form and solved by Benders decomposition. The proposed method is tested on an IEEE 24-bus system to illustrate its effectiveness in managing M-DERA integration while ensuring operational economics and mitigating the overloading of transmission lines.

## Nomenclature

Major symbols used throughout the paper are defined here, while others are clarified as needed after their first appearance.

### A. Indices and Sets

| | |
|---|---|
| $a$ | Index of M-DERAs. |
| $d, \mathcal{D}_a^A$ | Index and set of DERs underneath M-DERA $a$. |
| $g/l$ | Index of regular generators/transmission lines. |
| $\mathcal{G}^s$ | Set of quick-start generators. |
| $h, h^f$ | Indices of hyperplane components (HPCs) and high frequency components (HFCs). |
| $i, \mathcal{I}$ | Index and set of historical DF records. |
| $j, j'$ | Indices of generalized Gaussian distribution components (GGDCs). |
| $k$ | Index of iterations of the expectation–maximization algorithm. |

### A. Indices and Sets

| | |
|---|---|
| $m$ | Index of segments of piecewise linearization. |
| $n$ | Index of nodes in the transmission network. |
| $s, \mathcal{S}$ | Index and set of DF scenarios. |
| $t, \tau$ | Indices of time slots in the UC model. |
| $r, \mathcal{R}$ | Index and set of random samples. |

### B. Parameters

| | |
|---|---|
| $b_{g,m}^G/k_{g,m}^G$ / $b_{a,m}^A, k_{a,m}^A$ | Intercept and slope of the piecewise linear cost function of segment $m$ of generator $g$/M-DERA $a$. |
| $\overline{F}_l$ | Capacity limit of transmission line $l$. |
| $L_{n,t}$ | Active power load on node $n$ at time $t$. |
| $\overline{P}_a^A/\underline{P}_a^A$ | Real power output limits of M-DERA $a$. |
| $\overline{P}_g^G/\underline{P}_g^G$ | Real power output limits of regular generator $g$. |
| $\boldsymbol{R}_r$ | The $r$-th random sample used in approximating the integral via the Monte Carlo method. |
| $\overline{R}_a^A/\underline{R}_a^A$ | Ramp-up/down limit of M-DERA $a$. |
| $\overline{R}_g^G/\underline{R}_g^G$ | Ramp-up/down limit of regular generator $g$. |
| $\overline{R}_{g,t}^{sr}/\overline{R}_{g,t}^{nr}$ | Upper limit on spinning reserve (SR)/ non-spinning reserve (NR) of generator $g$ at time $t$. |
| $\hat{R}_t^{sr}/\hat{R}_t^{nr}$ | System SR/NR requirement at time $t$. |
| $SF_{l,n}$ | SF describing the sensitivity of the power injection of node $n$ to power flow on line $l$. |
| $T$ | Total time slots in UC. |
| $T_g^{on}/T_g^{off}$ | Minimum on-/off-time of generator $g$. |
| $z_{i,j}$ | Indicator encoding data $i$'s membership to GGDC $j$. |
| $\beta_j, \boldsymbol{\mu}_j, \boldsymbol{\Sigma}_j$ | Shape parameter, mean vector, and covariance matrix of GGDC $j$. |
| $\boldsymbol{\xi}_j$ | Parameters of GGDC $j$, i.e., $\boldsymbol{\xi}_j = \{\beta_j, \boldsymbol{\mu}_j, \boldsymbol{\Sigma}_j\}$. |
| $\boldsymbol{\xi}$ | Complete set of parameters of all GGDCs. |
| $\pi_j$ | Weight of GGDC $j$. |
| $\pi_h^{HPC}/\pi_{h^f}^{HFC}$ | Weight factor of HPC $h$/HFC $h^f$. |

### C. Variables

| | |
|---|---|
| $DF_{a,d,t}$ | DF of DER $d$ in M-DERA $a$ at time $t$. |
| $p_{a,t}^A/p_{g,t}^G$ | Power output of M-DERA $a$/generator $g$ at time $t$. |
| $r_{g,t}^{sr}/r_{g,t}^{nr}$ | SR/NR of generator $g$ at time $t$. |
| $S_{l,a,t}$ | Sensitivity of the power injection from M-DERA $a$ to power flow on line $l$ at time $t$. |
| $x_{g,t}$ | On-off status of generator $g$ at the time $t$. |
| $x_{g,t}^{nr}$ | NR commitment of generator $g$ at time $t$. |

### D. Functions

| | |
|---|---|
| $b(\cdot)$ | Index of the node where an asset is located at. |
| $f(\boldsymbol{X}|\boldsymbol{\xi}_j)$ | Probability density function (PDF) of unbounded generalized Gaussian of bounded GGDC $j$. |
| $P(\boldsymbol{X}|\boldsymbol{\xi})$ | PDF of the BHMM. |
| $p_h^{HPC}(\boldsymbol{X}|\boldsymbol{\xi}_h)$ | PDF of HPC $h$. |
| $p(\boldsymbol{X}|\boldsymbol{\xi}_j)$ | PDF of bounded GGDC $j$. |
| $\Gamma(\cdot)$ | Gamma function. |
| $\Psi(\cdot)$ | Digamma function, i.e., $\Psi(x) = d \ln \Gamma(x)/dx = \Gamma'(x)/\Gamma(x)$. |
| $\Psi'(\cdot)$ | Trigamma function, i.e., $\Psi'(x) = d\Psi(x)/dx$. |

This work was supported in part by the US Department of Energy Advanced Grid Modeling Program under Grant number RD-470006-23.

W. Wang, Z. Shao, and L. Wu are with the Department of Electrical and Computer Engineering, Stevens Institute of Technology, Hoboken, NJ 07030 USA (e-mail: wwang110, zshao7, lei.wu@stevens.edu).

Y. Liu is with the Electrical Engineering College, Sichuan University, Chengdu, 610017 China.

B. Eldridge, A. Somani, and J. T. Holzer are with Pacific Northwest National Laboratory, Richland, WA 99354.

# I. INTRODUCTION

## A. DER Aggregation in Wholesale Markets

Motivated by the urgent need to alleviate the energy crisis and reduce environmental pollution, distributed energy resources (DERs) have been undergoing rapid development and have already formed a significant presence in the power system. With this, efficiently integrating and managing a large fleet of DERs has become increasingly critical. Supported by policies such as FERC Order 2222 [1], novel system reforms have been proposed to enable DERs to participate aggregately in the wholesale electricity markets as a coordinated entity [2].

Practically, aggregation strategies can ease the management of DERs when integrated into wholesale electricity markets at scale. Economic dispatch [3], energy management [4], control schemes [5], and flexibility issues [6] of DER aggregations (DERAs) have been explored, often in the form of microgrids or virtual power plants [2]. Specifically, considering the interaction of DERAs with the regional transmission organizations (RTOs), many studies focus on the role of DERAs in optimizing operation and bidding strategies through bilevel [7] and multi-layer [8] models. In these studies, RTOs typically play a secondary, passive role, while the potential challenges of DERAs to the operational security and economics of the transmission system [9] have not been adequately explored from the RTO's perspective.

From the RTO's perspective, one main research direction has been focusing on coordinating DERAs and the RTO via a bi-layer model to facilitate their integration into the wholesale market operation [10]. In this bi-layer model, DERAs are operated by distribution system operators (DSOs) in the lower layer and the RTO acts as the transmission system operator (TSO) in the upper layer. In [11] and [12], the DERAs participate in the day-ahead wholesale energy and balancing markets through TSO-DSO coordination. Reference [13] formulates the TSO-DSO coordination within a day-ahead unit commitment (DAUC) model to design daily coordination plans. Other studies focus on building aggregated characteristics of DERAs so that the RTO only needs to consider the aggregated performance of DERAs, rather than individual DERs. Based on this idea, a feasibility region method is explored in [14] to estimate active and reactive power flow limits at the DERA-RTO interface, to assist the RTO in delimiting feasible operation plans with DERAs.

## B. Needs and Challenges of Integrating M-DERAs

Most research on the operation of RTOs with DERAs focuses on DERAs connected to the RTO network at a single transmission node. Catalyzed by the increasing scale of DERs, DERAs connecting with the RTO network via multiple nodes have attracted increasing attention from the industry [9], which are referred to as multi-transmission-node DERAs (M-DERAs) in this paper. For example, California Independent System Operator (CAISO) has been managing DERAs spanning multiple pricing nodes [15], PJM already dispatches demand response resources across varying geographic areas and different pricing nodes [16], and Midcontinent Independent System Operator (MISO) allows M-DERAs for capacity accreditation purposes [17]. However, the single-node DERA integration methods could struggle to handle M-DERAs, as they lead to more complex coupling between the RTO and DERs through multiple interfaces [15], which poses challenges to extending these methods to M-DERAs [18].

Indeed, although the FERC and RTOs have indicated that M-DERAs could bring potential benefits in promoting market entry and participation, reducing transaction costs, assembling appropriately sized resources, and lowering the risk of underperformance [15], [19], integrating M-DERAs also presents new challenges to RTOs. One of the most concerning issues is managing the impact of M-DERAs on transmission constraints. Specifically, when an M-DERA represents multiple underneath DERs to participate in the RTO market, the impacts of these DERs on transmission flows will be quantified via a single parameter. That is, the nodal shift factors (SFs) or power transfer distribution factors (PTDFs) of underneath DERs will be aggregated as a single value, which poses challenges for the RTO in appropriately quantifying the actual impacts of DERs on transmission power flows [20]. As shown in Fig. 1 where an M-DERA aggregating DER1 and DER2 is connected to the RTO network via two nodes, the aggregated SFs against different load levels could vary noticeably (i.e., 0.5429 in load level 1 versus 0.44 in load level 2). That is, the aggregated sensitivity of the M-DERA to transmission lines relies on dynamic system operation states. Thus, properly aggregating SFs or PTDFs is vital for meeting transmission constraints while ensuring market efficiency.

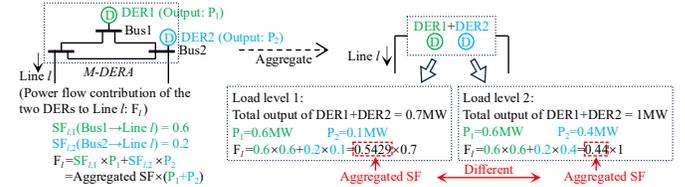

**Fig. 1.** An illustrative example of different aggregated SFs of an M-DERA under different loading levels.

To calculate aggregated sensitivity parameters, some research adopts partial network aggregation approaches [21]. In [22], a PTDF aggregation method for system congestion analysis is discussed; however, the congestion zone configuration highly depends on the system's operation state and is complicated for practical implementation. References [20] and [23]-[25] explore PTDF aggregation methods, which are independent of, or only slightly influenced by, nodal power injections, for transmission planning problems with acceptable accuracy. Nevertheless, these methods transform original transmission lines into equivalent pseudo connections between aggregated buses, thus discarding certain essential physical network characteristics [26], especially actual line capacity limits which are critical for ensuring operational security. Therefore, there remains a gap in calculating aggregated SFs or PTDFs of M-DERAs while still retaining an effective evaluation of physical transmission characteristics.

## C. Proposed Distribution Factor-Based M-DERA Integration

In industrial practice, the distribution factor (DF)-based M-DERA integration mechanism has gained momentum among RTOs and FERC [10]. DFs describe ratios of individual DERs' power contributions to the aggregated power output of the M-DERA.

Although DF offers an intuitive and concise approach for the RTO to describe the power share information of individual DERs underneath an M-DERA, it introduces new issues.

Specifically, the RTO must estimate the DF values (either by itself or by M-DERAs as a part of bidding parameters) to calculate the aggregated SFs of M-DERAs and clear the wholesale market; after that, each M-DERA self-dispatches its underneath DERs to follow RTO's aggregated dispatch instructions. However, DFs corresponding to each M-DERA's self-dispatch (SD) could deviate from the DFs used by the RTO, introducing uncertainties and leading to potential transmission overloading concerns for RTOs [15]. Indeed, from RTO's perspective, estimating DFs is a challenging task because: (i) DFs, depending on dispatches of individual DERs as well as the aggregated power of the M-DERA by definition, vary under different system conditions; and (ii) the SD process of M-DERAs is not transparent to RTOs. Thus, an appropriate DF estimation strategy is crucial for the RTO to satisfy transmission constraints while ensuring market efficiency.

With this, [9] and [18] analyze the impacts of different DF estimation strategies on the real-time market oscillation of MISO in real-time economic dispatch (RTED), signifying the importance of proper DF estimations. Indeed, as DAUC is an upper-stream decision process that feeds generator schedule plans to RTED, the influence of DF estimates used in DAUC will be further propagated to RTED, leading to more serious economics or transmission overloading concerns.

Fortunately, unlike RTED which has stringent computational time and cannot fully leverage historical DF records, in the DAUC, the RTO can use more historical DF records (i.e., based on state estimation results unveiling M-DERAs' SD results against RTO's aggregated instructions) and comprehensive techniques to capture DF dynamics under various operating conditions. For instance, one could fit historical DF records into certain multi-variant distribution functions to capture their uncertainties, which is then used to estimate future DFs. However, DF data usually does not conform to typical distributions because: (i) DFs are inherently bounded to be non-negative, which cannot be directly handled by common unbounded distributions; and (ii) driven by DERs' operational characteristics, DFs are usually concentrated on certain hyperplanes or multiple singletons (as detailed in later sections and Fig. 3). Such characteristics of DFs compromise the effectiveness of major fitting techniques.

In recognizing the insufficient exploration of DAUC with M-DERAs and the challenge of using typical distributions to describe DFs, we focus on the following contributions:

(i) DFs of M-DERAs are modeled as uncertain parameters in the DAUC via a bounded hetero-dimensional mixture model (BHMM). The BHMM is formulated as a weighted sum of high-frequency singletons and bounded multi-dimensional generalized Gaussian mixture model (BMGGMM) on multiple hyperplanes of different dimensions. The complete process of the BHMM, including grouping, dimensionality reduction, and fitting of historical DF records, is detailed in this paper. To the best of the authors' knowledge, this study is among the first to introduce this method to power system applications for fitting data distributed in a bounded hetero-dimensional space.

(ii) With uncertain DF parameters described by the BHMM, a chance-constrained unit commitment (CCUC) model integrated with M-DERAs is proposed, where transmission flow limits are formulated as chance constraints. The proposed CCUC model is further reformulated into a bi-level stochastic programming by sampling multiple DF scenarios out of the BHMM, and solved by Benders decomposition (BD).

The rest of this paper is organized as follows: Section II introduces the concept of DFs, the mechanism of the CCUC with M-DERAs, and the reformulation and solving methods; Section III presents the BHMM process, including grouping, dimension reduction, and fitting of historical DF records; The case studies are analyzed in Section IV to verify the feasibility and efficiency of the proposed DF-based SF aggregation strategy and the CCUC; The paper is concluded in Section V.

## II. Methodology of CCUC with M-DERAs

### A. The Formulation of DF-Based CCUC with M-DERAs

In this paper, an M-DERA is defined as a group of DERs that spans multiple transmission nodes within the RTO's network, as illustrated in Fig. 2. Each M-DERA participates in RTO operations as a single entity, featuring bidding price/ quantity, ramp rates, and capacity limits that submitted to the RTO to represent the collective characteristics of all underneath DERs. Once the aggregated dispatch instructions of M-DERAs are determined, each M-DERA executes the SD to determine the outputs of individual underneath DERs [18].

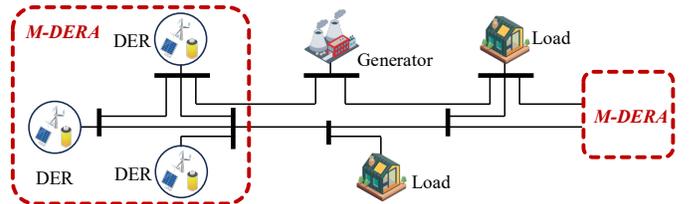

**Fig. 2.** The relationship between M-DERAs and individual DERs.

As each M-DERA acts as a single entity in the RTO market, we need to aggregate the sensitivities of transmission nodes underneath the M-DERA for properly assessing its contributions to transmission line flows while respecting the original physical line capacity limits. The DF-based SF aggregation strategy used by MISO [9] [18] is adopted here. DFs of individual DERs underneath an M-DERA are calculated as in (1.1), where $p_{a,d,t}^{DER}$ represents power output of DER $d$ in M-DERA $a$ at time $t$. For instance, if an M-DERA consists of two DERs with outputs of 1.2 MW and 0.8 MW, their respective DFs will be 0.6 and 0.4. By introducing the DFs, the sensitivity of M-DERA $a$ to line $l$ can be calculated as (1.2), describing that the aggregated sensitivity of an M-DERA equals the summed products of SFs of the nodes where underneath DERs are located at and their corresponding DFs.

$$DF_{a,d,t} = p_{a,d,t}^{DER}/p_{a,t}^{A}; \qquad \forall a, \forall d \in \mathcal{D}_a^A, \forall t \quad (1.1)$$

$$S_{l,a,t} = \sum_{d \in \mathcal{D}_a^A} DF_{a,d,t} \cdot SF_{l,b(d)} \qquad \forall a, \forall l, \forall t \quad (1.2)$$

Theoretically, DFs are tightly correlated with the M-DERA's SD process against RTO's aggregated instructions, which could be regarded as uncertain responses of M-DERAs from the RTO's perspective. With this, the RTO can develop a CCUC model to incorporate uncertain responses of M-DERAs through uncertain DF simulations and further ensure power flow limits via chance constraints. Moreover, as historical DF records shall consistently align with the M-DERA's SD process which pursues economic efficiency against RTO's aggregated instructions, it is reasonable to build proper probability distributions of DFs out of historical DF records for representing their uncertainties in CCUC. The construction of uncertain DF distributions will be detailed in Section III.

With established uncertain DF distributions, the proposed CCUC model is built to minimize the total daily operation cost, subject to prevailing physical and operation constraints of individual resources, system load balance, and uncertain DF-based chance constraints on transmission flow limits. The formulation is detailed in Appendix. Specifically, compared to the classic network-constraint UC (NCUC) model [27], the main difference lies in the DF-based chance constraints on transmission flow limits as specified in (2.1), which ensures that the probability of power flows remaining within line capacities is no less than a preset tolerance level $(1 - \varepsilon)$. The relationship between aggregated sensitivities of M-DERAs $S_{l,a,t}$ and uncertain DFs $\widehat{DF}_{a,d,t}$ is outlined in equation (2.2).

$$\mathbb{P}\{-\overline{F}_l \leq \sum_a S_{l,a,t} \cdot p_{a,t}^A + \sum_g SF_{l,b(g)} \cdot p_{g,t}^G - \sum_n SF_{l,n} \cdot L_{n,t} \leq \overline{F}_l; \forall l, \forall t\} \geq 1 - \varepsilon \tag{2.1}$$

$$S_{l,a,t} = \sum_{d \in \mathcal{D}_a^A} \widehat{DF}_{a,d,t} \cdot SF_{l,b(d)}; \qquad \forall a, \forall l, \forall t \tag{2.2}$$

The proposed CCUC is computationally challenged by two issues. First, the inverse cumulative density function of uncertain DFs, as presented in Section III, is typically not computable in closed form, such that the chance constraints (2.1) cannot be directly converted into equivalent deterministic reformulations. Second, the bilinear term $S_{l,a,t} \cdot p_{a,t}^A$ in (2.1), containing uncertain parameters and decision variables, poses significant computational challenges. To address these two issues, stochastic reformulation and BD method [28] are applied to reformulate (2.1) into a tractable deterministic form, as detailed in Section II.B.

### B. Stochastic Reformulation and Benders Decomposition

The fundamental principle of the stochastic reformulation and BD method involves discretizing the bilinear term into a stochastic formation via multiple scenarios, reformulating the resulting model into a linear mixed-integer programming (MIP), and solving it via an iterative BD process [28]. For the sake of discussion, the CCUC is compactly presented as in (3), where $\mathbf{x}$ are binary variables indicating generators' ON/OFF plans, $\mathbf{y}^G/\mathbf{y}^A$ are dispatches of generators/M-DERAs, and all other symbols are coefficient matrices and vectors, while $\widetilde{\mathbf{H}}^A$ involves uncertain DFs. Constraint (3.2) represents the chance constraint (2.1), and (3.3) encompasses all other constraints.

$$\min \mathbf{cx} + \mathbf{f}^A \mathbf{y}^A + \mathbf{f}^G \mathbf{y}^G \tag{3.1}$$

$$\text{s.t.} \quad \mathbb{P}\{\mathbf{Gx} + \widetilde{\mathbf{H}}^A \mathbf{y}^A + \mathbf{H}^G \mathbf{y}^G \geq \mathbf{h}\} \geq 1 - \varepsilon \tag{3.2}$$

$$\mathbf{Px} + \mathbf{Q}^A \mathbf{y}^A + \mathbf{Q}^G \mathbf{y}^G \geq \mathbf{r} \tag{3.3}$$

$$\mathbf{x} \in \{0,1\}^n, \mathbf{y}^G \in \mathbb{R}^{mG}, \mathbf{y}^A \in \mathbb{R}^{mA} \tag{3.4}$$

With given probability distributions of $\widetilde{\mathbf{H}}^A$, we can generate multiple scenarios to transform (3.2) into a stochastic formulation (4). Specifically, the chance constraints (3.2) are converted into a stochastic form (4.2) that can be linearized by the McCormick linearization method [28]. The enforcement of constraints for scenario $s$ is controlled by a binary variable $z_s$. The costs of individual scenarios are calculated in (4.4), and their weighted sum is included in the objective function as the expected operation cost of activated scenarios. The chance constraint tolerance level is controlled by (4.5).

$$\theta^{CC} = \min \mathbf{cx} + \sum_s \omega_s \eta_s \tag{4.1}$$

$$\text{s.t.} \quad (\mathbf{Gx} + \mathbf{H}_s^A \mathbf{y}_s^A + \mathbf{H}^G \mathbf{y}_s^G - \mathbf{h}) \cdot (1 - z_s) \geq 0; \quad \forall s \tag{4.2}$$

$$\mathbf{Px} + \mathbf{Q}^A \mathbf{y}_s^A + \mathbf{Q}^G \mathbf{y}_s^G \geq \mathbf{r}; \qquad \forall s \tag{4.3}$$

$$\eta_s = (\mathbf{f}^A \mathbf{y}_s^A + \mathbf{f}^G \mathbf{y}_s^G) \cdot (1 - z_s); \qquad \forall s \tag{4.4}$$

$$\sum_s \omega_s z_s \leq \varepsilon \tag{4.5}$$

$$\mathbf{x} \in \{0,1\}^n, \mathbf{y}_s^G \in \mathbb{R}^{mG}, \mathbf{y}_s^A \in \mathbb{R}^{mA}, z_s \in \{0,1\}; \quad \forall s \tag{4.6}$$

Directly solving (4) containing numerous scenarios can still be computationally intractable. Thus, an iterative BD method is further applied. In iteration $k$, given the decisions of the first stage variable $\hat{\mathbf{x}}^k$ and the dual variable $\boldsymbol{\gamma}_s$ corresponding to the primal constraints involving $\mathbf{y}_s^G$ and $\mathbf{y}_s^A$, the dual subproblem $\theta_s^{SP}$ of scenario $s$ can be formulated as shown in (5).

$$\theta_s^{SP} = \max \begin{bmatrix} \mathbf{h} - \mathbf{G}\hat{\mathbf{x}}^k \\ \mathbf{r} - \mathbf{P}\hat{\mathbf{x}}^k \end{bmatrix}^T \cdot \boldsymbol{\gamma}_s \tag{5.1}$$

$$\text{s.t.} \quad \begin{bmatrix} \mathbf{H}_s^A & \mathbf{H}^G \\ \mathbf{Q}^A & \mathbf{Q}^G \end{bmatrix}^T \cdot \boldsymbol{\gamma}_s \leq [\mathbf{f}^A \ \mathbf{f}^G]^T; \quad \boldsymbol{\gamma}_s \geq 0 \tag{5.2}$$

In iteration $k$, by solving (5), extreme points $u_s^k \in \mathcal{U}_s$ or extreme rays $v_s^k \in \mathcal{V}_s$ for scenario $s$ can be obtained. Then, the corresponding optimality cuts or feasibility cuts can be added to the master problem as shown in (6). By linearizing the optimality constraints (6.2) and feasibility constraints (6.3) via the McCormick linearization method [28], the master problem (6) can be solved by commercial MILP solvers. The computational performances of (6) can be further improved via multiple enhancement strategies, such as stochastic programming-based initialization and small-M based initialization [28]. Consequently, the optimal solution can be obtained via the iterative process described in Algorithm I.

$$\theta^{MP} = \min \mathbf{cx} + \sum_s \omega_s \eta_s \tag{6.1}$$

$$\text{s.t.} \quad \begin{bmatrix} \mathbf{h} - \mathbf{G}\hat{\mathbf{x}}^k \\ \mathbf{r} - \mathbf{P}\hat{\mathbf{x}}^k \end{bmatrix}^T \cdot u_s^k (1 - z_s) \leq \eta_s; \quad \forall u_s^k \in \mathcal{U}_s, \forall s, \forall k \tag{6.2}$$

$$\begin{bmatrix} \mathbf{h} - \mathbf{G}\hat{\mathbf{x}}^k \\ \mathbf{r} - \mathbf{P}\hat{\mathbf{x}}^k \end{bmatrix}^T \cdot v_s^k (1 - z_s) \leq 0; \quad \forall v_s^k \in \mathcal{V}_s, \forall s, \forall k \tag{6.3}$$

$$\sum_s \omega_s z_s \leq \varepsilon \tag{6.4}$$

$$\mathbf{x} \in \{0,1\}^n; z_s \in \{0,1\} \qquad \forall s \tag{6.5}$$

---

**Algorithm I**: Benders Decomposition Method

Set $LB = -\infty$; $UB = \infty$; $k = 1$; $\mathcal{U}_s = \mathcal{V}_s = \emptyset, \forall s$.
**for** iteration $k$:
    Solve (6) to obtain solutions $(\hat{\mathbf{x}}^k, z_s^k, \forall s)$ and $\theta^{MP,k}$.
    Update $LB \leftarrow \theta^{MP,k}$.
    **for** all activated $s$ such that $z_s^k = 0$:
        Solve (5) and obtain $\theta_s^{SP,k}$ and $u_s^k$ or $v_s^k$.
        Update $\mathcal{U}_s = \mathcal{U}_s \cup \{u_s^k\}$ or $\mathcal{V}_s = \mathcal{V}_s \cup \{v_s^k\}$.
        Add new Benders cut (6.2) or (6.3) to (6).
    **end for**
    Solve (4) with $(\mathbf{x}^k, z_s^k, \forall s)$ and obtain $\theta^{CC,k}$.
    Update $UB \leftarrow \min\{UB, \theta^{CC,k}\}$.
    If $(UB - LB)/LB$ is no larger than a predefined threshold, terminate and output $\hat{\mathbf{x}}^k$ as the final solution; otherwise, $k = k + 1$ and continue the iteration.
**end for**

---

In the above CCUC solving process, the stochastic scenarios, sampled from DF distributions to capture their uncertainty characteristics, would critically influence the solution quality. However, it is challenging to derive accurate DF distributions. Fig. 3 shows an example of DF records for an M-DERA with 3 DERs, generated following [18]. Evidently, these DF records do not conform to classical distributions (e.g., Gaussian, Gamma, and Poisson) or typical mixture models (e.g., Gaussian mixture model). Indeed, the distribution of DFs for an M-DERA with $n$ DERs may not be properly described by a $n$-dimensional distribution since the DF records are confined to specific hyperplanes (HPs) of diverse lower dimensions, which compromises the goodness of fit of typical distribution models. To address this issue, a novel BHMM is proposed, as detailed in Section III.

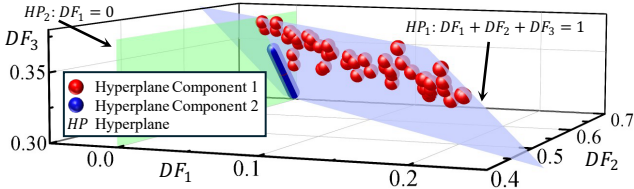

**Fig. 3.** An example of the DF records for an M-DERA with 3 DERs.

## III. BHMM FOR UNCERTAIN DFs

### A. Properties of DFs and DF Data Grouping

DFs present the following properties:

**Property 1**: The sum of all DFs of an M-DERA is equal to 1.

**Property 2**: Each DF value is bounded by [0,1].

**Property 3**: Certain DF data may occur with much higher frequency, due to physical and/or economic operation characteristics of M-DERAs.

These properties indicate that DFs of individual DERs in an M-DERA are linearly dependent, and some DF data reside on certain HPs of diverse lower dimensions. For example, all DF data in Fig. 3 lie on $HP_1$: $DF_1 + DF_2 + DF_3 = 1$, and the blue ones also lie on $HP_2$: $DF_1 = 0$. It suggests that for a certain subset of DF records, as DFs of some DERs are either fixed or can be derived from other DFs, we shall use a distinct probability function of proper dimension to better fit them. In Fig. 3, as $DF_1$ of all data on $HP_2$ are zero, the $DF_1$ dimension can be removed and $DF_2 + DF_3 = 1$ is sufficient to describe all data on $HP_2$; further, only one distribution of $DF_2$ or $DF_3$ needs to be built, and the other one can be derived according to $DF_2 + DF_3 = 1$. Consequently, the DF data in Fig. 3 can be split into two groups: one group contains red points that can be analyzed via a 2-D distribution, and the other group contains blue points that can be analyzed via a 1-D distribution.

In this paper, such groups are referred to as hyperplane components (HPCs) hereinafter. Additionally, according to Property 3, high-frequency DF data at singletons shall also be grouped separately to avoid over- or under-fitting when they are mixed with others. Such groups of high-frequency DFs are termed as high-frequency components (HFCs).

This grouping strategy can be generalized to DF data of any $D$-dimension. First, DF data whose frequency exceeds a predefined threshold are grouped into HFCs; then, the remaining data are put into multiple HPCs based on coincident zero DF elements. The theoretical maximum number of HPCs is $2^D$. The grouping process is summarized in Algorithm II.

The HPCs obtained by Algorithm II have reducible dimensions, i.e., the dimensions associated with coincident zero DFs in each HPC can be reduced. One more dimension can be further reduced according to the linear dependency of DFs in Property 1. For the example in Fig. 3, $DF_3$ of all data in HPC2 (blue) are zero, and $DF_1$ or $DF_2$ can be further reduced because $DF_1 + DF_2 = 1$; Likewise, in HPC1 (red), any one of the three DFs can be reduced.

---

## Algorithm II : DF Data Grouping Process

Input DF data $\mathcal{I} = \{X_i, i = 1 \dots l\}$, where each $X_i$ is a $D$-dimension DF record of $D$ DERs in an M-DERA. Count the number of occurrences of each DF record $X_i$ in $\mathcal{I}$ as $O_i$.

Set $\mathcal{I}_h = \emptyset, h = 0 \dots (2^D - 1)$ and high-frequency threshold $f$.

**for** $i = 1 \dots l$:

   **if** $O_i \geq f$, Assign all such identical DF records to one HFC and remove them from $\mathcal{I}$.

**end for**

**for** $h = 0 \dots (2^D - 1)$

   Convert $h$ to a binary representation, identify all of its zero digits, and collect their indices to set $\mathcal{D}^*$.

   **for** all remining DF data $X_i \in \mathcal{I}$:

      **if** $X_{i,d} = 0, \forall d \in \mathcal{D}^*$, **then** $\mathcal{I}_h = \mathcal{I}_h \cup \{X_i\}, \mathcal{I} = \mathcal{I} \backslash \{X_i\}$.

   **end for**

**end for** when the $h = (2^D - 1)$ or $\mathcal{I} = \emptyset$.

---

By grouping DF data into HFCs and HPCs and further reducing their dimensionalities, the overall DF "distribution" can be represented by the weighted sum of probabilities of HFCs and the probability density functions (PDFs) of HPCs, as shown in (7). Weighting factors $\pi_{h^f}^{HFC} / \pi_h^{BC}$ describe the proportions of DF data of groups $h^f / h$ contained in the historical DF data, and, $p_{h^f}^{HFC}(X_{h^f}) / p_h^{HPC}(X_h|\xi_h)$ is the probability/PDF of HFC $h^f$/HPC $h$. It is crucial to clarify that $P(X|\xi)$ in BHMM is not a typical PDF in the conventional sense, as it includes non-zero probabilities at certain singletons in HFCs. Nonetheless, the BHMM can still reasonably describe data with specific spatial characteristics that are often tricky to capture in general full-dimensional distributions.

$$P(X|\xi) = \sum_{h^f} \pi_{h^f}^{HFC} \cdot p_{h^f}^{HFC}(X_{h^f}) + \sum_h \pi_h^{HPC} \cdot p_h^{HPC}(X|\xi_h) \quad (7)$$

### B. Fitting HPCs: BMGGMM

In each HPC, the DF data are scattered within bounded ranges of certain dimensions, i.e., they do not conform to classic unbounded distributions. Thus, BMGGMM is implemented to capture the distribution pattern of DF data within a bounded region. BMGGMM is an extension of GMM to overcome its shortcomings, such as the rigidity of shape and limited performance, in many real applications within bounded regions [29]. A 1-D illustrative example of a HPC bounded within [0,0.7] is shown in Fig. 4 to compare the performance of classic distributions and BMGGMM over a bounded region. Fig. 4 illustrates that BMGGMM can capture both the probability pattern within the range and the zero probability out of the range, while others have evident errors.

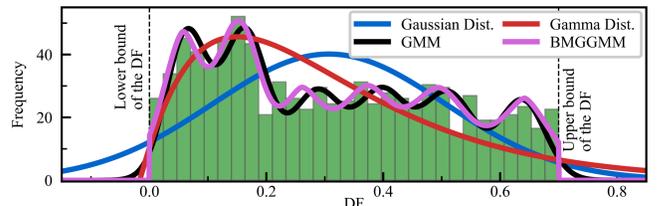

**Fig. 4.** An illustrative example showing the performance of classic distributions and BMGGMM for a HPC within a bounded region.

The PDF of BMGGMM for each HPC can be presented as the weighted sum of multivariate generalized Gaussian distribution components (GGDCs) $p(X|\xi_j)$ defined within a bounded region $\Omega$, as shown in (8.1), where $f(X|\xi_j)$ is the PDF of unbounded multivariate GGDC as shown in (8.2) [30]. For simplicity, the subscript $h$ for HPCs is omitted in this part.

$$p^{HPC}(X|\xi) = \sum_j \pi_j \cdot p(X|\xi_j) = \sum_j \pi_j \cdot \frac{f(X|\xi_j)}{\int_\Omega f(u|\xi_j)du} \quad (8.1)$$

$$f(X|\mu_j, \Sigma_j, \beta_j) = \left\{ \Gamma\left(\frac{D}{2}\right) \Big/ \left[ \pi^{\frac{D}{2}} \cdot \Gamma\left(\frac{D}{2\beta_j}\right) \cdot 2^{\frac{D}{2\beta_j}} \right] \right\} \cdot \left( \beta_j / |\Sigma_j|^{\frac{1}{2}} \right) \cdot$$
$$\exp\left\{ -\frac{1}{2} \cdot \left[ (X - \mu_j)^T \cdot \Sigma_j^{-1} \cdot (X - \mu_j) \right]^{\beta_j} \right\}; \quad \forall j \quad (8.2)$$

To estimate the parameters $\mu_j$, $\Sigma_j$, $\beta_j$, and $\pi_j$ in (8.1) and (8.2), the maximum likelihood approach is implemented via the expectation-maximization (EM) algorithm. Specifically, the log-likelihood of the PDF of BMGGMM can be given as (8.3) [29], where posterior probabilities $z_{i,j}$ is defined in (8.4).

$$\mathcal{L}(\mathcal{X}, \mathcal{Z}, \xi) = \sum_i \sum_j \left\{ \ln\left[ \pi_j \cdot p(X|\xi_j) \right] \right\}^{z_{i,j}} \quad (8.3)$$

$$z_{i,j} = \left[ \pi_j \cdot p(X_i|\xi_j) \right] \Big/ \left[ \sum_{j'} \pi_{j'} \cdot p(X_i|\xi_{j'}) \right]; \quad \forall i, \forall j \quad (8.4)$$

Then the parameters can be estimated by calculating the partial deviations of (8.3) with respect to $\mu_j$, $\Sigma_j$, and $\beta_j$, which are detailed in the following subsections. It is noteworthy that EM is an iterative process, and in the following discussions, only the parameter being updated at the current iteration ($k+1$) is annotated by superscript ($k+1$), while other parameters are set as solutions from iteration $k$.

### 1). Estimation of Means

The estimation of means $\mu_j$ is based on the fixed-point method [31]. Specifically, for each GGDC $j$, by reorganizing $\partial \mathcal{L} / \partial \mu_j = 0$ in the form of $\mu_j^{(k+1)} = g_{\mu_j}(\mu_j^{(k)})$, $\mu_j^{(k+1)}$ can be updated by the $\mu_j$ solution from iteration $k$ as in (9.1)-(9.3).

$$\mu_j^{(k+1)} = \sum_i z_{i,j} \cdot \left[ y_{i,j}^{\beta_j-1} X_i - \frac{\int_\Omega f(u|\xi_j) \cdot (y_j^{(u)})^{\beta_j-1} \cdot (u - \mu_j^{(k)}) du}{\int_\Omega f(u|\xi_j) du} \right] \Big/ \sum_i z_{i,j} \quad (9.1)$$

$$y_{i,j} = (X_i - \mu_j)^T \cdot \Sigma_j^{-1} \cdot (X_i - \mu_j) \quad \forall i \quad (9.2)$$

$$y_j^{(u)} = (u - \mu_j)^T \cdot \Sigma_j^{-1} \cdot (u - \mu_j) \quad (9.3)$$

The two integrals in (9.1) can be tackled via numerical methods, such as the Monte Carlo method. Denote $|\mathcal{R}|$ random samples in $\Omega$ as $R_r$, $r \in \mathcal{R}$, the two integrals in (9.1) can be approximated as shown in (9.4)-(9.6).

$$\int_\Omega f(u|\xi_j) \cdot (y_j^{(u)})^{\beta_j-1} \cdot (u - \mu_j) du$$
$$\approx \frac{1}{|\mathcal{R}|} \sum_r f(R_r|\xi_j) \cdot (y_j^{(R_r)})^{\beta_j-1} \cdot (R_r - \mu_j) \quad (9.4)$$

$$y_j^{(R_r)} = (R_r - \mu_j)^T \cdot \Sigma_j^{-1} \cdot (R_r - \mu_j) \quad \forall r \quad (9.5)$$

$$\int_\Omega f(u|\xi_j) du \approx \frac{1}{|\mathcal{R}|} \sum_r f(R_r|\xi_j) \quad (9.6)$$

Finally, $\mu_j^{(k+1)}$ can be calculated by (9.7).

$$\mu_j^{(k+1)} \approx \sum_i z_{i,j} \cdot \left[ y_{i,j}^{\beta_j-1} \cdot X_i - \frac{\sum_r f(R_r|\xi_j) \cdot (y_j^{(R_r)})^{\beta_j-1} \cdot (R_r - \mu_j^{(k)})}{\sum_r f(R_r|\xi_j)} \right] \Big/ \sum_i z_{i,j} \quad (9.7)$$

### 2). Estimation of Covariance

The estimation of the covariance matrix $\Sigma_j$ can be obtained via a similar process. Specifically, for each GGDC $j$, by reorganizing $\partial \mathcal{L} / \partial \Sigma_j = 0$ in the form of $\Sigma_j^{(k+1)} = g_{\Sigma_j}(\Sigma_j^{(k)})$, the covariance $\Sigma_j$ at iteration ($k+1$) can be updated as shown in (10.1). By further applying the similar idea as in (9.4)-(9.6), $\Sigma_j^{(k+1)}$ can be calculated as shown in (10.2).

### 3). Estimation of Shape Parameter

The shape parameter $\beta_j$ is calculated using the Newton method. Specifically, $\beta_j$ is the root of $\partial \mathcal{L} / \partial \beta_j = 0$ and can be iteratively solved via (11.1)-(11.5). The integrals in (11) can be handled via the same numerical methods discussed above.

$$\beta_j^{(k+1)} = \beta_j - \frac{\partial \mathcal{L}}{\partial \beta_j} \cdot \left( \frac{\partial^2 \mathcal{L}}{\partial \beta_j^2} \right)^{-1} \quad (11.1)$$

$$\frac{\partial \mathcal{L}}{\partial \beta_j} = \sum_i z_{i,j} \left\{ Q_j^{(1)} - \frac{1}{2} \cdot y_{i,j}^{\beta_j} \cdot \ln y_{i,j} - \frac{\int_\Omega f(u|\xi_j) \cdot \left[ Q_j^{(1)} - \frac{1}{2}(y_j^{(u)})^{\beta_j} \cdot \ln y_j^{(u)} \right] du}{\int_\Omega f(u|\xi_j) du} \right\} \quad (11.2)$$

$$Q_j^{(1)} = \frac{1}{\beta_j} + \Psi\left(\frac{D}{2 \cdot \beta_j}\right) \cdot \frac{D}{2 \cdot \beta_j^2} + \ln 2 \cdot \frac{D}{2 \cdot \beta_j^2} \quad (11.4)$$

$$Q_j^{(2)} = -\frac{1}{\beta_j^2} - \Psi'\left(\frac{D}{2 \cdot \beta_j}\right) \cdot \left(\frac{D}{2 \cdot \beta_j^2}\right)^2 - \Psi\left(\frac{D}{2 \cdot \beta_j}\right) \cdot \frac{D}{\beta_j^3} - \ln 2 \cdot \frac{D}{\beta_j^3} \quad (11.5)$$

### 4). Estimation of Weights

The weights $\pi_j$, satisfying $\pi_j \geq 0, \forall j$, and $\sum_j \pi_j = 1$, can be iteratively estimated by (12), where $N$ is the number of data.

$$\pi_j^{(k+1)} = \sum_i z_{i,j}/N \quad (12)$$

The complete EM algorithm for estimating the parameters of the BMGGMM is summarized in Algorithm III.

---

**Algorithm III**: EM Algorithm for BMGGMM

Input observation data $\mathcal{X}$.

Set $k = 1$. For $\forall j$, initialize $\mu_j^k$, $\Sigma_j^k$, and $\pi_j^k$ via regular GMM, set $\beta_j^k = 1$. Calculate the initial log-likelihood by (8.3). Set coverage threshold $\epsilon$.

**if** the change in log-likelihood $\geq \epsilon$

  **E-step:**

    Update posterior probabilities $z_{i,j}^k$ by (8.4).

  **M-step:**

    Update $\mu_j^{(k+1)}$, $\Sigma_j^{(k+1)}$, $\beta_j^{(k+1)}$, and $\pi_j^{(k+1)}$ by (9.7), (10.2), (11.1), and (12).

    Update the log-likelihood by (8.3), and set $k = k + 1$.

**end if**. Return final estimations of $\mu_j^k$, $\Sigma_j^k$, $\beta_j^k$, and $\pi_j^k$.

---

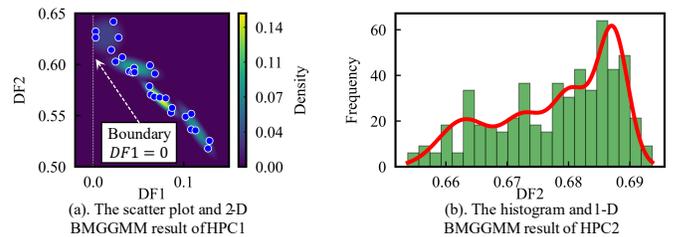

(a). The scatter plot and 2-D BMGGMM result of HPC1

(b). The histogram and 1-D BMGGMM result of HPC2

**Fig. 5.** An illustrative example showing the PDFs of DFs of HPC1 and HPC2.

$$\Sigma_j^{(k+1)} = \frac{1}{\sum_i z_{i,j}} \cdot \sum_i z_{i,j} \cdot \left\{ \beta_j \cdot y_{i,j}^{\beta_j-1} \cdot (X_i - \mu_j) \cdot (X_i - \mu_j)^T - \left[ \int_\Omega f(u|\xi_j) \cdot \left[ \beta_j \cdot (y_j^{(u)})^{\beta_j-1} \cdot (u - \mu_j) \cdot (u - \mu_j)^T - \Sigma_j \right] du \right] \Big/ \int_\Omega f(u|\xi_j) du \right\} \quad (10.1)$$

$$\Sigma_j^{(k+1)} \approx \frac{1}{\sum_i z_{i,j}} \cdot \sum_i z_{i,j} \cdot \left\{ \beta_j \cdot y_{i,j}^{\beta_j-1} \cdot (X_i - \mu_j) \cdot (X_i - \mu_j)^T - \sum_r \left[ f(R_r|\xi_j) \cdot \left[ \beta_j \cdot (y_j^{(R_r)})^{\beta_j-1} \cdot (R_r - \mu_j) \cdot (R_r - \mu_j)^T - \Sigma_j \right] \right] \Big/ \sum_r f(R_r|\xi_j) \right\} \quad (10.2)$$

$$\frac{\partial^2 \mathcal{L}}{\partial \beta_j^2} = \sum_i z_{i,j} \left\{ Q_j^{(2)} - \frac{1}{2} \cdot (y_{i,j})^{\beta_j} \cdot (\ln y_{i,j})^2 - \frac{\int_\Omega \left\{ f(u|\xi_j) \cdot \left[ Q_j^{(1)} - \frac{1}{2}(y_j^{(u)})^{\beta_j} \cdot \ln y_j^{(u)} \right]^2 \right\} du}{\int_\Omega f(u|\xi_j) du} - \frac{\int_\Omega f(u|\xi_j) \cdot \left[ Q_j^{(2)} - \frac{1}{2}(y_j^{(u)})^{\beta_j} \cdot (\ln y_j^{(u)})^2 \right] du}{\int_\Omega f(u|\xi_j) du} + \frac{\left\{ \int_\Omega f(u|\xi_j) \cdot \left[ Q_j^{(1)} - \frac{1}{2}(y_j^{(u)})^{\beta_j} \cdot \ln y_j^{(u)} \right] du \right\}^2}{\left[ \int_\Omega f(u|\xi_j) du \right]^2} \right\} \quad (11.3)$$

By applying Algorithm III, PDFs of individual HPCs can be properly generated over the respective bounded regions. PDFs of HPC1 and HPC2 in Fig. 3 are illustrated in Fig. 5. These PDFs will be used to generate the DF scenarios via a typical rejection sampling process [32] to solve the CCUC.

## V. CASE STUDY

In this section, the proposed CCUC model is tested on a modified IEEE 24-bus case [33], as shown in Fig. 6. It includes 24 nodes (N1-N24), 12 generators (G1-G12), 34 transmission lines (L1-L34), and one M-DERA which aggregates 4 DERs (D1-D4) spanning over N11, N12, N19, and N20. The target day covers 24 hours (H1-H24) for DAUC and 288 5-minute time slots (M1-M288) for RTED implementation. The load curve is sourced from PJM [34]. Other system data are described in [35].

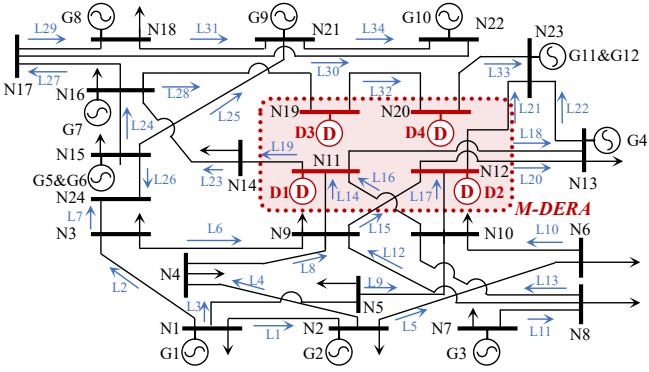

**Fig. 6.** The network of the modified IEEE 24-bus case.

### A. Comparison of Different DF Estimation Strategies

To demonstrate the effectiveness of the proposed DF estimation strategy, the following three cases are tested:

• *C1: Uniform DF Case*, which fixes the DFs of all DERs as the reciprocal of the number of DERs in the M-DERA (i.e. 0.25 in this case) and solves the NCUC model;

• *C2: Historical DF Case*, which sets the DFs as the average of historical DFs records and solves the NCUC model;

• *C3: BHMM-based DF Case*, which fits DF distributions by BHMM and solves the CCUC model with the chance constraint tolerance level $\varepsilon$ of 5%.

In C2 and C3, historical operation records of the past 7 days are used to generate DF records and fit DF distributions. The results of BHMM in C3 are demonstrated in Table I.

TABLE I RESULTS OF BHMM IN C3

| HFC&HPC | Dimension | Number of components | Proportion |
|---|---|---|---|
| HFCs | 1 | 21 HFCs | 66.54% in total |
| HPC1 | 2 | 5 GGDCs | 5.28% |
| HPC2 | 3 | 3 GGDCs | 1.35% |
| HPC3 | 3 | 5 GGDCs | 3.49% |
| HPC4 | 3 | 5 GGDCs | 2.75% |
| HPC5 | 4 | 5 GGDCs | 20.24% |

The effects of generator ON/OFF plans under different cases are further evaluated by implementing these plans in a 5-min rolling RTED to first determine the aggregated dispatch instruction of the M-DERA [18]; then, given the aggregated dispatch instruction, the following three M-DERA SD strategies are applied to compare the ultimate consequence of ON/OFF plans derived by different DF estimation strategies:

• *SD1*: the M-DERA sorely minimizes the cost of underneath

DERs regardless of impacts on transmission power flows [18], i.e., following (13.1) and (13.2);

• *SD2*: the M-DERA minimizes the cost of underneath DERs while considering potential congestion on transmission lines, i.e., following (13.1)-(13.4);

• *SD3*: the M-DERA strictly follows the DFs in the RTED i.e., following (13.1), (13.5), and $\Delta p_{l,a,t} = 0, \forall l$.

Note that (13) is processed in rolling for each time $t$. The superscript "*" signifies that a variable is fixed based on the UC solution. $F^*_{l,a,t}$ represents the power flow of M-DERA $a$ on line $l$ at time $t$ calculated in UC, and $\Delta p_{l,a,t}$ is a slackness variable with a penalty cost $C^P(\Delta p_{l,a,t})$ in objective (13.1).

$$\min \sum_d C^D(p^{DER}_{d,t}) + \sum_l C^P(\Delta p_{l,a,t}) \quad (13.1)$$
$$\text{s.t.} \quad p^{A*}_{a,t} = \sum_{d \in D^A_a} p^{DER}_{d,t} \quad (13.2)$$
$$\left| \sum_d SF_{l,b(d)} \cdot p^{DER}_{d,t} \right| \leq |F^*_{l,a,t}| + \Delta p_{l,a,t}; \quad \forall l \quad (13.3)$$
$$\Delta p_{l,a,t} \geq 0; \quad \forall d \quad (13.4)$$
$$p^{DER}_{d,t} = p^{A*}_{a,t} \cdot DF^*_{d,t}; \quad \forall l \quad (13.5)$$

Table II reports the results of C1-C3 against the three SDs. It is observed that C3 has the lowest total UC cost, which is also revealed in the RTED against all three SDs. In terms of the generation costs under the same SD, C3 is the lowest, averaging 4.9% and 5.9% less than C1 and C2, respectively. C1 and C2 also undergo various levels of load shedding (LS) to maintain system power balance in RTED, while C3 has no LS across all SDs. The M-DERA cost after SD depends on specific SD strategies, where C3 does not have a clear advantage. Nevertheless, C3 exhibits the lowest total cost (i.e., the sum of generation cost and LS cost of RTED as well as SD cost of the M-DERA) under all three SDs, averaging 4.9% and 6.2% less than C1 and C2, respectively. Table II demonstrates that under different SDs, C3 outperforms C1 and C2 in both total costs and LS, indicating that the proposed DF estimate strategy can derive more effective ON/OFF plans for systems with M-DERAs under various RTED and SD conditions.

TABLE II COST COMPARISON OF DIFFERENT CASES AND SD STRATEGIES (K$)

| Case | Total UC Cost | SD | RTED Costs | | M-DERA SD Cost | Total Cost |
|---|---|---|---|---|---|---|
| | | | Generation Cost | LS Cost | | |
| C1 | 2,075.04 | SD1 | 1,965.65 | 5.34 | 15.16 | 1,986.15 |
| | | SD2 | 1,804.81 | 5.34 | 156.75 | 1,966.91 |
| | | SD3 | 1,680.59 | 5.34 | 257.34 | 1,940.86 |
| C2 | 2,045.87 | SD1 | 1,987.11 | 14.26 | 22.19 | 2,023.57 |
| | | SD2 | 1,834.16 | 14.26 | 157.14 | 2,005.56 |
| | | SD3 | 1,686.94 | 0 | 262.59 | 1,949.53 |
| C3 | **1,951.29** | SD1 | **1,870.50** | **0** | 16.98 | **1,887.47** |
| | | SD2 | **1,710.15** | **0** | 158.21 | **1,868.36** |
| | | SD3 | **1,603.40** | **0** | 248.56 | **1,851.96** |

Fig. 7 highlights the differences in generator ON/OFF plans under C1-C3, which occur among G3, G5, G8, G9, and G11 and mainly during H1 to H9. Specifically, in C1 and C2, G9 is off from H1 to H9; G8 is off for 4 hours in C1 and 3 hours in C2; G3, G5, and G11 are off to varying extents in C2. In contrast, in C3, G3 starts to be online at H8 and G11 remains online all the time. The most significant differences among C1-C3 are G3 and G9, which are indeed caused by different DF estimation strategies. Table III shows SFs of G3, G9, and M-DERA to L28 and L12, which are two heavily loaded lines operating at 100% capacity in the negative direction. G9, G3, and M-DERA are the three most expensive generation assets in the system. From C1 to C2, a higher SF of M-DERA helps alleviate congestion on L12 and L28, allowing other cheaper generators to operate with lower total UC costs. Consequently, G3 shuts down from H1 to H9 in C2. Despite SF of the M-DERA in C3 further increases, the expensive G9 is switched on as its SF to L28 is much higher than G3 and M-DERA, enabling it to better congestions on L28 under multiple scenarios in CCUC.

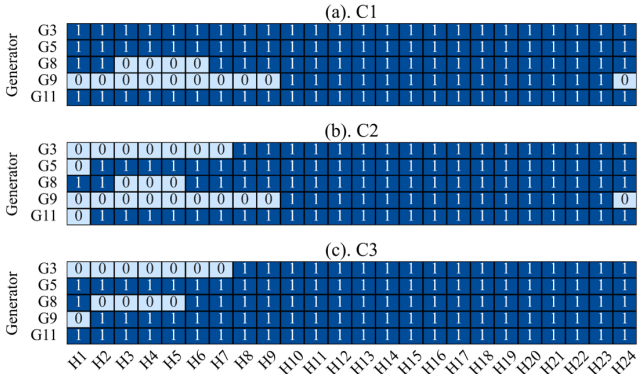

**Fig. 7.** Highlights on ON/OFF plan differences in the three cases.

TABLE III SFs IN THE UC UNDER DIFFERENT DF ESTIMATION STRATEGIES

| Line | G3 | G9 | M-DERA | | |
| | | | C1 | C2 | C3 (Average SF of activated scenarios) |
|---|---|---|---|---|---|
| L12 | 0.5054 | -0.0087 | -6.43e-4 | 2.98e-5 | 1.52e-4 |
| L28 | 0.1687 | 0.7270 | -0.0073 | 0.0190 | 0.0228 |

The difference in ON/OFF plans leads to different capabilities of the system to respond to load changes in RTED with various LS costs. The LS in RTED is caused by different capabilities of generators in handling power differences between hourly UC and 5-min RTED. Specifically, LS in C1 and C2 happens at N16 at H6, H9, and H24, during which G9 is offline in C1 and C2 but online in C3. G3 and G9 are two of the fastest ramping generators in the system. In C2, the absence of G3 and G9 leads to insufficient ramping capability, thus impairing the ability to handle real-time load fluctuations and leading to the highest LS costs. In C1, although G3 is online, its long electrical distance to N16 prevents it from supplying power to N16 through many heavily loaded lines. Thus, LS of C1, though lower than C2 under SD1 and SD2, still exists. In C3, the presence of G9 and its short distance to N16 offer a stronger capability to handle real-time load changes. As a result, no LS occurs under all SDs in C3. Such results highlight that the proposed DF strategy can derive more suitable unit ON/OFF plans for systems with M-DERA, leading to superior economic and less LS in RTED.

In terms of transmission power flows after SD, in all cases, all transmission lines are operated within the limits except for L28. The loading level on L28 throughout the day is further demonstrated in Fig. 8, which is defined as the ratio of power flow after SD to line capacity. In Fig. 8, each subplot divides the day into two intervals based on the similarity of loading levels. In Interval A, under each SD, C1 always faces the most critical overloading, suffering multiple overloading in SD1 and the longest durations at full capacity in SD2 and SD3; C2 does not experience overloading under all SDs but still encounters full-load operation; C3 sees no overloading and negligible full-load operations. The comparison of L28's loading levels in Interval A under three DF strategies validates that C3 can maintain a proper margin for critical lines to handle potential line overloading caused by SDs of M-DERAs.

In Interval B, overloading situations for three DF strategies under the same SD are similar but vary across different SDs. Specifically, overloading under SD1 is the most severe as the M-DERA executes SD1 regardless of line limits, potentially leading to the largest deviation from the DFs in RTED and the most severe overloading. In SD2, overloading is alleviated because line limits are considered in the SD. Nevertheless, overloading still occurs at certain time slots as the M-DERA does not have sufficient capacities to adjust dispatches of DERs under the given aggregated dispatch instruction while ensuring the safety of transmission lines. In SD3, as the M-DERA strictly follows the DFs in RTED, the actual impacts of the DERs on transmission lines are identical to the estimates in RTED, and no overloading occurs. Such observations validate the significance of properly estimating DFs for effectively integrating M-DERAs into RTO operations.

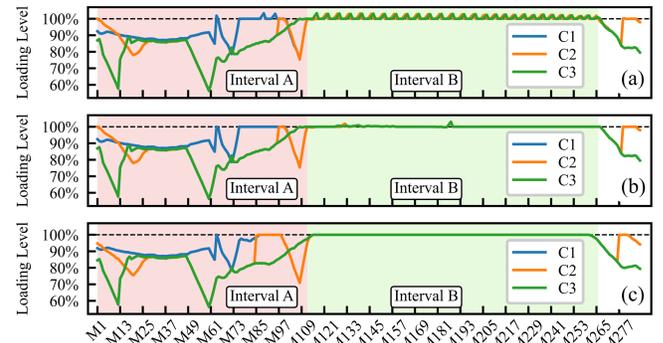

**Fig. 8.** The loading levels on L28 under SD1 (a), SD2 (b), and SD3 (c).

### B. Sensitivity Test of CCUC

This subsection analyzes the impact of different settings of $\varepsilon$ in CCUC using the same implementation process as in Subsection IV.A, and the costs are compared in Table IV.

Table IV COST COMPARISON OF DIFFERENT $\varepsilon$ AND SD STRATEGIES (k$)

| $\varepsilon$ | Total UC Cost | SD | RTED Costs | | M-DERA SD Cost | Total Cost |
| | | | Generation Cost | LS Cost | | |
|---|---|---|---|---|---|---|
| 0% | 2,061.35 | SD1 | 1,965.65 | 5.34 | 15.16 | 1,986.15 |
| | | SD2 | 1,804.81 | 5.34 | 156.75 | 1,966.91 |
| | | SD3 | 1,680.59 | 2.93 | 257.34 | 1,940.86 |
| 2.5% | 1,958.62 | SD1 | 2,042.67 | 3.64 | 14.57 | 2,060.88 |
| | | SD2 | 1,880.95 | 3.64 | 156.43 | 2,041.02 |
| | | SD3 | 1,757.00 | 6.20 | 257.27 | 2,020.47 |
| 5% | 1,951.29 | SD1 | 1,870.50 | 0 | 16.98 | 1,887.47 |
| | | SD2 | 1,710.15 | 0 | 158.21 | 1,868.36 |
| | | SD3 | 1,603.40 | 0 | 248.56 | 1,851.96 |
| 7.5% | 1,911.08 | SD1 | 1,861.66 | 0 | 17.38 | 1,879.04 |
| | | SD2 | 1,698.38 | 0 | 161.07 | 1,859.44 |
| | | SD3 | 1,589.22 | 0 | 254.39 | 1,843.61 |

In terms of total UC cost, the results align with the general conclusions of chance constraints, i.e., the lower the tolerance level $\varepsilon$, the higher the cost to satisfy the chance constraint requirements. As shown in Table IV, the total UC cost is the highest when $\varepsilon = 0\%$ because the capacity constraints on transmission lines are strictly enforced. As the $\varepsilon$ increases, more DF scenarios are considered non-mandatory, reducing the number of scenarios that the system must meet for transmission limits, and accordingly, the UC costs decrease.

Under the same $\varepsilon$, costs vary across SDs. From SD1 to SD3, both total cost and RTED generation costs drop, while the M-DERA SD cost gradually rises. These trends are consistent with those in Table I, reaffirming the impact of operation strategies of RTED with M-DERAs on system economics.

Comparing the results of different $\varepsilon$ under the same SD yields different observations. Although the total UC cost decreases with lower $\varepsilon$, RTED costs do not follow the same trend. As detailed in Table IV, under the same SD, the total RTED cost with $\varepsilon = 2.5\%$ is the highest, followed by 0%, 5%, and 7.5%. The generation cost follows a similar trend, while the M-DERA costs under three SDs at various $\varepsilon$ do not show a consistent trend. The LS costs also do not necessarily decrease against the increase in $\varepsilon$. In SD1 and SD2, the LS costs with $\varepsilon = 0\%$ is higher than that with $\varepsilon = 2.5\%$; while in SD3, such relationship is reversed. When $\varepsilon = 5\%$ and 7.5%, no LS occurred under any of the three SDs.

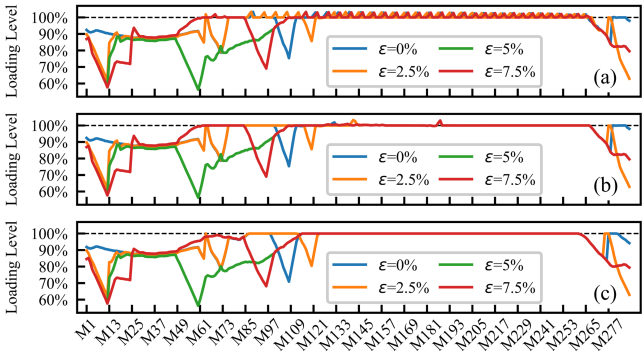

**Fig. 9.** The loading levels on L28 under different $\varepsilon$ and SD1 (a), SD2 (b), and SD3 (c).

Table V L28 Overloading under Different $\varepsilon$ and SD Strategies

| SD | $\varepsilon$ | Number of Overloading Time Slots | Average Overloading Level |
|---|---|---|---|
| SD1 | 0% | 110 | 1.06% |
| | 2.5% | 108 | 1.14% |
| | 5% | 88 | 1.20% |
| | 7.5% | 101 | 1.11% |
| SD2 | 0% | 42 | 0.37% |
| | 2.5% | 27 | 0.58% |
| | 5% | 31 | 0.50% |
| | 7.5% | 38 | 0.40% |
| SD3 | All $\varepsilon$ | 0 | 0 |

The loading levels on L28 under different $\varepsilon$ and SDs are further illustrated in Fig. 9 and Table V. Overloading trends from Fig. 9 are consistent with Fig. 8, i.e., SD1 experiences the most severe overloading, and SD3 the least. Under the same SD, the longest overloading time under SD1 and SD3 both occurs in the 0% case, sequentially followed by 2.5%, 7.5%, and 5%. However, in SD2, the 2.5% case experiences the shortest overloading time. Although the 0% case has the longest overloading time across all SDs, the average overloading level, calculated as the total loading levels over the overloading duration, is the smallest across all cases.

Observations from Table IV, Table V, and Fig. 9 also suggest that ON/OFF plans with a smaller $\varepsilon$ are not always derive better performance for specific RTED and/or M-DERA implementations. First, in the proposed CCUC with uncertain DFs, ON/OFF plans with a smaller $\varepsilon$ are derived to handle a collective of scenarios with higher probabilities. Thus, when the actual RTED and/or SD of M-DERAs deviate from these scenarios, the ON/OFF plans may not perform the best. Second, implementing hourly ON/OFF plans to minute-wise RTED may lead to a natural mismatch if the minute-wise system conditions (such as loads) are not adequately reflected in the CCUC. Thus, the research on coordinating DAUC and RTED involving M-DERAs deserves further investigation.

## VI. Conclusion

Inspired by industrial practices and FERC Order 2222's DERA integration requirements, this paper introduces a DF-based SF aggregation strategy within the UC framework. In this approach, from the RTO's perspective, DFs are treated as uncertainty parameters in UC, with transmission capacity limits formulated as chance constraints. The derived CCUC-based method includes some computational challenges that are addressed by a scenarios-based stochastic form reformulation and a BD-based solution method. To effectively capture DFs on hetero-dimensional hyperplanes, a BHMM is designed to group data, reduce dimensionality, and accurately fit the DF data onto the designated hyperplanes using the BMGGMM.

The proposed CCUC with the BHMM-based DF estimation is evaluated by applying ON/OFF plans in the rolling RTED and subsequent SD of M-DERAs. Comparisons against other DF strategies reveal that the BMGGMM-based DF strategy can derive more effective ON/OFF plans with augmented economic (reducing UC, RTED, and total costs across multiple SDs) and safety (mitigating RTED LS and transmission overloading after SD) benefits. Further, the sensitivity test reveals that the performance of the proposed DF strategy is influenced by the hierarchical DAUC-RTED-SD process, highlighting the importance of further research on coordinating the proposed CCUC and the downstream applications to further promote the effective integration of M-DERAs. Implementing the proposed CCUC would also need to resolve issues associated with scenario-based constraints and market settlements. Stochastic-based modeling in RTOs creates well-known market integration issues, which may be solved by removing the stochastic elements from market clearing [36] or through modeling approximations [37], [38]. We will leverage these works to explore the market integration issues of the proposed CCUC model in future work.

## Appendix

The formulation of the CCUC is presented in (A.1)-(A.22). The objective function (A.1) minimizes the total operation cost, including the start-up and shut-down costs of regular generators, as well as the operation costs of regular generators and M-DERAs as in (A.2) and (A.3).

Operational characteristics of individual generators are described in (A.4)-(A.14), including the on/off switching logic (A.4) and (A.5), the min on- and off-time requirements (A.6) and (A.7), the limits on power outputs and spinning/non-spinning reserves (A.8)-(A.12), and the ramp-up and -down limits (A.13) and (A.14). Constraints (A.15) and (A.16) describe power output and ramping constraints of M-DERAs.

Constraints (A.17)-(A.19) describe the system power balance and SR/NR requirements. Constraint (A.20) is the chance-constrained transmission flow limits, and (A.21) calculates the aggregated sensitivity of each M-DERA to lines via SFs and DFs. The integrality requirements of binary variables are specified in (A.22).

$$\min \sum_t \left[ \sum_g \left( C^{ON} \cdot u_{g,t} + C^{OFF} \cdot v_{g,t} + C_{g,t} \right) + \sum_a C_{a,t} \right] \quad \text{(A.1)}$$

$$\text{s.t.} \quad C_{g,t} \geq k_{g,m}^G \cdot p_{g,t}^G + b_{g,m}^G \cdot x_{g,t}; \qquad \forall g, \forall m, \forall t \quad \text{(A.2)}$$

$$C_a \geq k_{a,m}^A \cdot p_{a,t}^A + b_{a,m}^A; \qquad \forall a, \forall m, \forall t \quad \text{(A.3)}$$

$$u_{g,t} \geq x_{g,t} - x_{g,t-1}; \qquad \forall g, \forall t \quad \text{(A.4)}$$

$$v_{g,t} \geq x_{g,t-1} - x_{g,t}; \qquad \forall g, \forall t \quad \text{(A.5)}$$

$$x_{g,t} - x_{g,t-1} \leq x_{g,\tau};$$
$$\forall g, \forall t, \forall \tau \in \left[t+1, \min\{t+T_g^{on}-1, T\}\right] \quad \text{(A.6)}$$

$$x_{g,t-1} - x_{g,t} \leq 1 - x_{g,\tau};$$
$$\forall g, \forall t, \forall \tau \in \left[t+1, \min\{t+T_g^{off}-1, T\}\right] \quad \text{(A.7)}$$

$$p_{g,t}^G - r_{g,t}^{sr} \geq \underline{P_g^G} \cdot x_{g,t}; \qquad \forall g, \forall t \quad \text{(A.8)}$$

$$p_{g,t}^G + r_{g,t}^{sr} \leq \overline{P_g^G} \cdot x_{g,t}; \qquad \forall g, \forall t \quad \text{(A.9)}$$

$$0 \leq r_{g,t}^{sr} \leq \overline{R_g^{sr}} \cdot x_{g,t}; \qquad \forall g, \forall t \quad \text{(A.10)}$$

$$\underline{P_g^G} \cdot x_{g,t}^{nr} \leq r_{g,t}^{nr} \leq \overline{R_g} \cdot x_{g,t}^{or}; \qquad \forall g \in \mathcal{G}^s, \forall t \quad \text{(A.11)}$$

$$x_{g,t}^{or} + x_{g,t} \leq 1; \qquad \forall g \in \mathcal{G}^s, \forall t \quad \text{(A.12)}$$

$$p_{g,t}^G - p_{g,t-1}^G \leq \overline{R_g^G} \cdot x_{g,t-1} + \underline{P_g^G} \cdot (x_{g,t} - x_{g,t-1})$$
$$+ \overline{P_g^G} \cdot (1 - x_{g,t}); \qquad \forall g, \forall t \quad \text{(A.13)}$$

$$p_{g,t-1}^G - p_{g,t}^G \leq \overline{R_g^G} \cdot x_{g,t} + \underline{P_g^G} \cdot (x_{g,t-1} - x_{g,t})$$
$$+ \overline{P_g^G} \cdot (1 - x_{g,t-1}); \qquad \forall g, \forall t \quad \text{(A.14)}$$

$$0 \leq p_{a,t}^A \leq \overline{P_a^A}; \qquad \forall a, \forall t \quad \text{(A.15)}$$

$$-\underline{R_a^A} \leq p_{a,t}^A - p_{a,t-1}^A \leq \overline{R_a^A}; \qquad \forall a, \forall t \quad \text{(A.16)}$$

$$\sum_a p_{a,t}^A + \sum_g p_{g,t}^G = \sum_n L_{n,t}; \qquad \forall t \quad \text{(A.17)}$$

$$\sum_g r_{g,t}^{sr} \geq \hat{R}_t^{sr}; \qquad \forall t \quad \text{(A.18)}$$

$$\sum_g (r_{g,t}^{sr} + r_{g,t}^{nr}) \geq \hat{R}_t^{sr} + \hat{R}_t^{nr}; \qquad \forall t \quad \text{(A.19)}$$

$$\mathbb{P}\{-\overline{F_l} \leq \sum_a S_{l,a,t} \cdot p_{a,t}^A + \sum_g SF_{l,b(g)} \cdot p_{g,t}^G -$$
$$\sum_n SF_{l,n} \cdot L_{n,t} \leq \overline{F_l}; \forall l, \forall t\} \geq 1 - \varepsilon \quad \textcolor{red}{\text{(A.20)}}$$

$$S_{l,a,t} = \sum_{d \in \mathcal{D}_a^A} \widetilde{DF}_{a,d,t} \cdot SF_{l,b(d)}; \qquad \forall a, \forall l, \forall t \quad \text{(A.21)}$$

$$u_{g,t}, v_{g,t}, x_{g,t}, x_{g,t}^{or} \in \{0,1\} \qquad \forall g, \forall t \quad \text{(A.22)}$$